\documentclass[aip,apl,amsmath,amssymb,reprint]{revtex4-1}

\usepackage{graphicx}% Include figure files
\usepackage{dcolumn}%
\usepackage{bm}%
\usepackage[acronym,shortcuts]{glossaries}

\makeglossaries
\newacronym{mce}{MCE}{magnetocaloric effect}
\begin{document}

%\preprint{}

\title{History dependence of directly observed magnetocaloric effects in~(Mn,~Fe)As} %Title of paper



\author{Milan Bratko}
\email[]{m.bratko09@imperial.ac.uk}
%\thanks{}
\affiliation{Department of Physics, Blackett Laboratory, Imperial College London, London, SW7 2AZ, United Kingdom}

\author{Kelly Morrison}
\affiliation{Department of Physics, Blackett Laboratory, Imperial College London, London, SW7 2AZ, United Kingdom}

\author{Ariana de Campos}
\affiliation{Instituto de Ci\^{e}ncias Tecnol\'{o}gicas e Exatas, Universidade Federal do Tri\^{a}ngulo Mineiro - UFTM, Uberaba, MG, Brazil}

\author{Sergio Gama}
\affiliation{Departamento de Ci\^{e}ncias Exatas e da Terra, Universidade Federal de S\~{a}o Paulo, Campus Diadema, Diadema, SP, Brazil}

\author{Lesley F. Cohen}
\affiliation{Department of Physics, Blackett Laboratory, Imperial College London, London, SW7 2AZ, United Kingdom}

\author{Karl G. Sandeman}
\affiliation{Department of Physics, Blackett Laboratory, Imperial College London, London, SW7 2AZ, United Kingdom}


\date{\today}

\begin{abstract}
We use a calorimetric technique operating in sweeping magnetic field to study the thermomagnetic history-dependence of the magnetocaloric effect (MCE) in Mn$_{0.985}$Fe$_{0.015}$As. We study the magnetization history for which a ``colossal'' MCE has been reported when inferred indirectly via a Maxwell relation. We observe no colossal effect in the direct calorimetric measurement. We further examine the impact of mixed-phase state on the MCE and show that the first order contribution scales linearly with the phase fraction. This validates various phase-fraction based methods developed to remove the colossal peak anomaly from Maxwell-based estimates.
\end{abstract}


\maketitle %\maketitle must follow title, authors, abstract and \pacs


The first order nature of the Curie transition in MnAs has been of interest for many years and was the motivation behind Bean and Rodbell's well-known model of first order magnetic phase transitions \cite{Bean:1962}. Recently MnAs has been in the spotlight again after the report of a ``colossal'' \ac{mce}---an order of magnitude larger than usual \cite{Gama:2004}. It, along with anomalous peaks in the \ac{mce} of other systems, was inferred indirectly from magnetization measurements via a Maxwell relation and disputed because of a comparison with calorimetric measurements showing no colossal effect \cite{Liu:2007, Tocado:2009}. The calorimetric measurements, however, follow a significantly different thermomagnetic history. Here we use a calorimetric technique operating in a field-sweeping mode to reproduce several thermomagnetic histories and thus investigate the possibility of observing an \ac{mce} that depends on the specific magnetization history.

The \ac{mce} can be quantified as an adiabatic temperature change, $\Delta T_{ad}$, or an isothermal entropy change, $\Delta S$, for a certain magnetic field variation. The latter can be estimated via a Maxwell relation:
\begin{equation}
\left(\frac{\partial S}{\partial B}\right)_{T}=\left(\frac{\partial M}{\partial T}\right)_{B},
\label{eq:max}
\end{equation}
where $M$ is the magnetization and $B$ is the magnetic flux density, which we later assume to be equal to $\mu_{0}H$. The use of this relation requires a knowledge of magnetization as a function of $B$ and $T$ and it is usually obtained from isothermal magnetization measurements taken in discrete temperature steps. It is important to note that the Maxwell relation is based on equilibrium thermodynamics, i.e. assuming that magnetization and entropy are single-valued functions of state. This is true in materials with a continuous phase transition, however, first order phase transitions usually display finite hysteresis due to the presence of multiple local minima in the free energy separated by an energy barrier. This enables the existance of metastable, non-equilibrium states. Nevertheless, the Maxwell relation is often used, assuming that the effect of a small and finite region of irreversibility is negligible.

A ``colossal'' \ac{mce} was first reported in MnAs under hydrostatic pressure \cite{Gama:2004} followed by its observation in Mn$_{1-x}$Fe$_{x}$As\cite{Campos:2006} and Mn$_{1-x}$Cu$_{x}$As\cite{Rocco:2007} at ambient pressure. The effect was later reported even in MnAs at ambient pressure \cite{Ranke:2006}, in disagreement with previous reports \cite{Wada:2001, Gama:2004}. The authors in Ref.~\onlinecite{Ranke:2006} claimed the capture of a colossal effect was possible due to smaller temperature steps between the magnetization isotherms. Interestingly, the temperature step was significantly smaller than the thermal hysteresis of the Curie transition, indicating a breakdown of the assumption that the irreversibility is small \cite{Caron:2009} and that use of the Maxwell relation is questionable.

The common feature of all the reports of colossal \ac{mce} is the use of an indirect approach involving isothermal magnetization measurements and the Maxwell relation given in Eq.~\ref{eq:max}. In the temperature range of the colossal effect the magnetization isotherms display plateaus of intermediate saturation before the metamagnetic transition occurs suggesting a mixed-phase state (e.g. Fig. 3b in Ref.~\onlinecite{Campos:2006}) and the first order phase transition displays a rather large thermal/field hysteresis.

A preferred method for measuring the \ac{mce} is calorimetry, where more relevant thermodynamic variables (heat capacity or even the heat flux associated with the \ac{mce}) can be measured directly. The claimed colossal effect was disputed because of a comparison with such measurements where no colossal effect was observed \cite{Liu:2007, Tocado:2009}. Various numerical and graphical approaches have been proposed to remove the spurious colossal peak from the entropy change estimates \cite{Liu:2007, Tocado:2009, Balli:2009, Cui:2010, Das:2010}. It has been shown that a non-colossal \ac{mce} estimate can be obtained directly from isothermal magnetization data by following a particular magnetization protocol, known as a `loop process' \cite{Caron:2009}, where the sample is reset to a well defined low-field state before every remagnetization.  

Nevertheless, it is important to note that the aforementioned calorimetric measurements were all performed in a temperature-sweeping mode (isofield measurement). The Maxwell estimate from isofield magnetization measurements also shows no colossal \ac{mce} \cite{Wada:2001}. So far, the magnetization protocol that was claimed to lead to colossal effect via the Maxwell relation, has not been reproduced in a direct calorimetric measurement. Wang \emph{et al.} \cite{Wang:2010} have compared Maxwell estimates with calorimetric measurements taken in field-sweeping mode, however, while they follow the usual thermal protocol in the magnetization measurements, they adopt the loop process in the calorimetric measurements. Here we reproduce the exact magnetization history that led to a reported colossal \ac{mce} via the Maxwell relation,  but instead carry out a direct calorimetric measurement. We thereby conduct a first examination of the effects of thermomagnetic history on directly observed \ac{mce} including the possibility of observing the colossal effect. We also examine the impact that a mixed-phase state would have on the \ac{mce}. 

We measured a sample from the same series of Mn$_{1-x}$Fe$_{x}$As compounds studied in Ref.~\onlinecite{Campos:2006}. As in MnAs, Fe-substituted samples in the series undergo a first order magnetostructural phase transtion from a hexagonal ferromagnetic (FM) state to an orthorhombic paramagnetic (PM) state. Here we study the $x=0.015$ composition (sample preparation as described in Ref.~\onlinecite{Campos:2006}). We note that the sample displays a virgin effect. In order to induce a FM ground state, an 8~T field was applied at 220~K \cite{Fjellvag:1988}. After the initial treatment, the sample had a reproducible Curie transition at 280~K on heating ($T_{C}^{heating}$) and at 255~K on cooling ($T_{C}^{cooling}$). This particular sample was chosen to match the temperature range of our calorimeter, limited to below 295~K.

We first studied two magnetization histories which we call History A and History B. In History A we applied and removed the magnetic field (8~T) isothermally at temperatures starting from 295~K down to 265~K in 2.5~K steps. After each isothermal measurement we cooled the sample directly to the next measurement temperature. In History B  we performed the isothermal measurements from 295~K down to 230~K in steps of 5~K, however, after every magnetization measurement the sample was reset to the PM state by heating it to 295~K before cooling to the next isothermal measurement temperature (in zero field). History B is equivalent to the loop process described in Ref.~\onlinecite{Caron:2009} used to recover a non-colossal estimate of the isothermal entropy change from the Maxwell relation.

\begin{figure}
\includegraphics{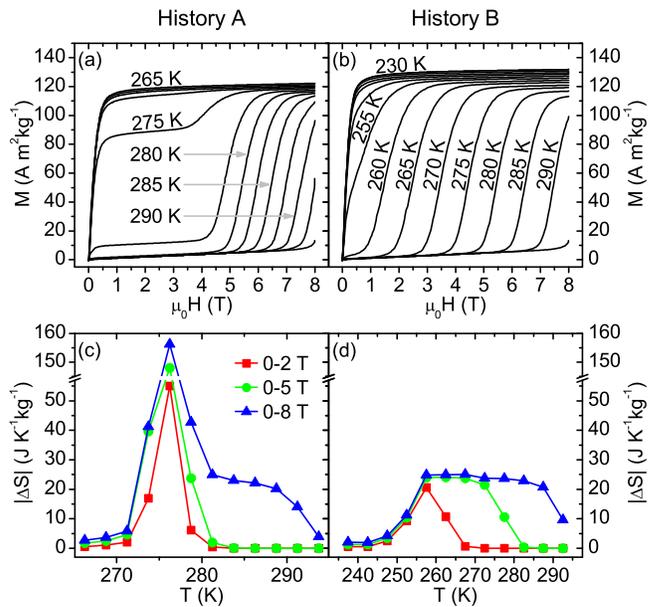}
\caption{\label{fig:mag} Bulk magnetometry data on field application (a,~b) and isothermal entropy changes (c,~d) as estimated by the Maxwell relation for History A and B, respectively. History A leads to the colossal estimate of MCE, as indicated by the peak at $\sim$275~K.}
\end{figure}

All magnetization measurements were performed in a Quantum Design PPMS. The magnetization data for both histories are presented in Fig.~\ref{fig:mag}a~and~\ref{fig:mag}b. History~A features the characteristic plateaus around $T_{C}^{heating}$ and use of the Maxwell relation results in the colossal \ac{mce} estimate in this temperature range (Fig.~\ref{fig:mag}c). In History B, the entropy change concentrated in the colossal peak for History A is distributed across the whole range of thermal hysteresis from $T_{C}^{cooling}$ to $T_{C}^{heating}$, yielding a non-colossal estimate of the \ac{mce}.

We compare this result with the calorimetric measurements, which were obtained using a SiN membrane-based microcalorimeter operating in a field-sweeping mode \cite{Minakov:2005, Miyoshi:2008}. The technique measures small fragments of the order of $\mu$g. We used the same fragment (`fragment 1') to reproduce History A and History B and note that calibration of the measurement is very sensitive to sample position and orientation. Because of the violent nature of the first order magnetostructural phase transition, the sample had to be fixed to the SiN membrane with GE varnish and the sample mass could not be measured afterwards. The specific values were obtained by comparison with another, pre-weighed fragment (`fragment 2').

Importantly, the instrument operates in two distinct setups measuring either the latent heat of a first order phase transition or the heat capacity. Thus we repeated both magnetization histories (`A' and `B') for the separate measurement of the latent heat and the heat capacity.

The latent heat, $Q_{LH}$, is measured in a `quasi-adiabatic temperature probe' setup \cite{Miyoshi:2008}. It is measured as a change in local temperature. In strongly first order materials (such as the one studied here), this is registered as one or more sharp spikes in a thermopile voltage as the sample is field-driven through the first order phase transition. These spikes decay back to the bath temperature with a time constant of approximately 1~s. The thermopile voltage is converted to heat by comparison to a calibration heat pulse provided by a local heater. We interpret this directly as a first order contribution to the total isothermal entropy change:
\begin{equation}
\Delta S_{LH}(T)=-\frac{Q_{LH}(T,H_C)}{T}.
\label{eq:LH}
\end{equation}
Because of the short time constant and relatively slow field ramp (0.5~T/min), continuous changes of entropy (second order effects) are picked up only as fluctuations in the baseline of the thermopile voltage.

We therefore recover the second order contribution to the total entropy change instead from heat capacity data measured in the `ac heat capacity probe' setup \cite{Minakov:2005}. Here the use of an ac temperature modulation combined with a lock-in measurement ($\tau$~=~1~s) suppresses the detection of latent heat \cite{Morrison:2011} and we measure only the heat capacity as a function of magnetic field. For evaluation of the second order contribution to the isothermal entropy change, temperature dependence of heat capacity is necessary:
\begin{equation}
\Delta S_{HC}(T)=\Delta S(T_{ref})+\int_{T_{ref}}^{T}\frac{\Delta C_p(T^\prime)}{T^\prime}\,\mathrm{d}T^\prime
\label{eq:HC}
\end{equation}
where $\Delta S(T_{ref})$ is a reference point obtained from magnetization measurements, well away from the first order phase transition. 

To avoid any complications associated with heat capacity data within the integral crossing the first order phase transition \cite{Morrison:2012} we consider the change in heat capacity of the PM phase between zero applied field and the critical field, $H_C$, at the chosen temperature, $T$, and integrate it from a high temperature reference point down to $T$ (provided that the sample was in the PM state at the beginning of the measurement). For a second order contribution from the FM state above $H_C$ we consider the change of heat capacity of the FM state between $H_C$ and the maximum applied field and integrate it from a low temperature reference point up to $T$.

The total entropy change on field application (the sum of Eqs.~\ref{eq:LH}~and~\ref{eq:HC}) is dominated by latent heat (over 90\%), as expected for a material with a strong first order phase transition. Any second order contributions come from the FM state only. Within instrument resolution, the heat capacity of the PM state appears to be field-independent. The Maxwell relation shows negligible entropy changes in the PM state (see Fig.~\ref{fig:mag}d at high temperatures for low field variations where the sample remains in the PM state over the whole field range).

\begin{figure}
\includegraphics{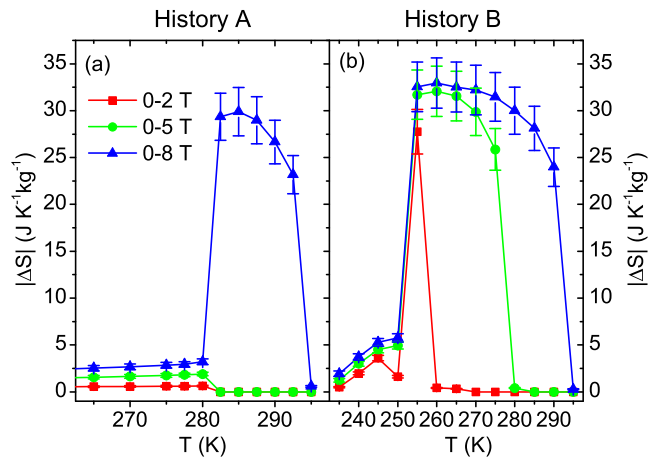}
\caption{\label{fig:cal} Total isothermal entropy change on field application as observed by microcalorimetry on fragment 1. History A (a) shows non-colossal MCE and simply restricts the large entropy change to a narrower temperature range when compared with History B (b).}
\end{figure}

The resulting total entropy changes for both histories are shown in Fig.~\ref{fig:cal}. We observe no colossal effect in the direct measurement. History A simply restricts the entropy change to a temperature range above $T_{C}^{heating}$ where the phase transition is reversible upon field removal. Thus, we show that the apparent entropy change concentration around $T_{C}^{heating}$ in History A is artificial.

Nevertheless, there is a sudden drop of entropy change from 282.5~K to 280.0~K---no latent heat is observed at the lower temperature. This indicates that while the sample starts from a mostly PM state at 282.5~K, it remains trapped in a fully FM state upon field removal at this temperature and no latent heat is observed on the next field application at 280.0~K. Thus, at the chosen measurement temperatures, no significant mixed-phase state was induced in the sample. 

In case one may suggest that the colossal entropy change is necessarily associated with the mixed-phase state (as it is a common feature in the magnetization measurements in the temperature range with colossal \ac{mce} estimates), we deliberately study a third magnetization history where we vary the mixed-phase fraction.

We performed a measurement on the pre-weighed fragment 2 at 295~K, at which temperature the PM state is well-defined at 0~T. Due to the slightly lower critical fields of this fragment, we observed one dominant latent heat spike just below 8~T even at 295~K. To vary the mixed-phase fraction we utilized the fact that the reverse transition occurs in a cascade of smaller spikes. At first we observed the latent heat upon field application (just below 8~T) and removal (at $\sim$3~T). We then set the mixed-phase state on the second---partial---field removal where we stopped the field removal roughly in the middle of the cascade (3.27~T). We determine the fraction of the FM state converted back to the PM state by comparing the latent heat with the previous (full) field removal. In the subsequent field application back to the maximum field we drove the sample through the first order phase transition in the mixed-phase state.

\begin{table}
\caption{\label{tab:LH}Latent heat component of isothermal entropy change in a mixed-phase state at 295~K (fragment 2).}
\begin{ruledtabular}
\begin{tabular}{lrr}
\textbf{Step} & \boldmath${\Delta S_{LH}}$ & \textbf{Relevant ratio}\\
\textbf & \textbf{[J K$^{-1}$ kg$^{-1}$]} & \textbf{[\%]}\\
1: 0 to 8 T & -16.7 $\pm$ 1.1\\
2: 8 to 0 T & +13.5 $\pm$ 0.9\\
3: 0 to 8 T & -16.4 $\pm$ 1.1\\
4: 8 to 3.27 T & +8.2 $\pm$ 0.5& 61 $\pm$ 3 (of 2)\\
5: 3.27 to 8 T & -10.0 $\pm$ 0.7 & 61 $\pm$ 3 (of 3)\\
6: 8 to 0 T & +13.2$\pm$ 0.9
\end{tabular}
\end{ruledtabular}
\end{table}

The measurement is summarised in Table~\ref{tab:LH}. In step~4 we set a mixed-phase state by reverting only 61\% of the FM phase induced in step~3. Upon the subsequent field application (step~5) we observe 61\% of the latent heat as compared with the full field application in step~3. Thus, the latent heat on field application scales simply with the PM phase fraction. We observe no colossal latent heat (entropy change) when inducing the phase transition in a mixed-phase state.

Note that there is a large difference in the latent heat on field application and removal in a full magnetization cycle (e.g. steps~1 and~2 in Table~\ref{tab:LH}). This is a consequence of the large hysteresis and will be the subject of future work. The reproducibility of latent heat in steps~1 and~3 confirms that the transition is fully reversed in step~2. For phase fraction determination we always consider the change in the latent heat relative to a comparable process (e.g. partial field removal with respect to full field removal).

In conclusion, we have shown that while there are differences in the directly observed \ac{mce} due to thermomagnetic history, they do not amount to a colossal \ac{mce}. We thus confirm that the concentration of entropy change resulting in the colossal \ac{mce}, as given by the Maxwell relation, is artificial. We also show that colossal \ac{mce} cannot be observed by driving the sample through the phase transition in a mixed-phase state. We confirm quantitatively that the first order contribution to the total entropy change scales with the phase fractions \cite{Kuepferling:2008}. This validates various numerical and graphical approaches used to recover non-colossal \ac{mce} estimates from the magnetization data using phase fractions, e.g. in Refs.~\onlinecite{Liu:2007, Tocado:2009, Balli:2009, Cui:2010, Das:2010}. Future work will use the large hysteresis in MnAs to examine the effect of hysteresis on out-of-equilibrium \ac{mce} in a magnetic field cycle, since this has been discussed mainly theoretically until now \cite{Amaral:2009}.

% If in two-column mode, this environment will change to single-column format so that long equations can be displayed. 
%\begin{widetext}
%$$\mbox{put long equation here}$$
%\end{widetext}

% Figures should be put into the text as floats. 
% Use the graphics or graphicx packages (distributed with LaTeX2e).
% Put the label that you will use with \ref{} command in the braces of the \label{} command.
%
% \begin{figure}
% \end{figure}

% \begin{table}
% \end{tabular}
% \end{table}

\begin{acknowledgments}
The authors thank A.D. Caplin for useful discussions and gratefully acknowledge financial support from: The Royal Society; the EPSRC through Grant EP/G060940/1; FAPEMIG through Grant Nos. APQ-01367-10, APQ-04163-10 and APQ-01037-11; CNPq through Grant Nos. 308181/2009-3 and 474431/2010-0; and FAPESP---Funda\c{c}\~{a}o de Amparo \`{a} Pesquisa do Estado de S\~{a}o Paulo through Grant No. 01/0558-3.
\end{acknowledgments}

%\bibliography{myLibrary}
%

\end{document}